\documentclass[conference,compsoc]{IEEEtran}

\ifCLASSOPTIONcompsoc
  \usepackage[nocompress]{cite}
\else
  \usepackage{cite}
\fi

\ifCLASSINFOpdf
\else
\fi

\usepackage{url}
\usepackage{xurl}

\usepackage[nolist]{acronym}
\usepackage{amsmath} 
\usepackage{xspace}
\usepackage{siunitx}
\usepackage{amssymb}
\usepackage{todonotes}
\setuptodonotes{inline}
\usepackage{graphicx}
\usepackage{subcaption}
\usepackage{tabularx}
\usepackage{booktabs}
\usepackage{multirow}
\usepackage{hyperref}

\newcommand{\cf}{\textit{cf}.\xspace}
\newcommand{\etal}[0]{\textit{et~al.}\xspace}
\newcommand{\eg}{\textit{e.g.}\xspace}
\newcommand{\ie}{\textit{i.e.}\xspace}
\newcommand{\Paragraph}[1]{\smallskip\noindent\textbf{#1}\hspace{0.5em}}
\usepackage{tcolorbox}
\newtcolorbox{takeawaybox}[1][]{colback=gray!10, colframe=black, boxrule=0.5pt, arc=2pt, boxsep=0pt, left=4pt, right=4pt, top=8pt, bottom=4.4pt, before skip=6pt, after skip=6pt, #1}
\newcommand{\takeaway}[2][]{\begin{takeawaybox}[#1]#2\ifhmode\strut\fi\end{takeawaybox}}

\newtheorem{definition}{Definition}

\begin{document}

\title{Spoofing Wireless Sensing with Metasurface-based Physical Deepfakes}

\author{\IEEEauthorblockN{Johannes Kortz\IEEEauthorrefmark{1},
Vignesh Natarajkumar\IEEEauthorrefmark{1}\IEEEauthorrefmark{2},
Christof Paar\IEEEauthorrefmark{1} and
Paul Staat\IEEEauthorrefmark{3}}

\IEEEauthorblockA{\IEEEauthorrefmark{1}Max Planck Institute for Security and Privacy, \{johannes.kortz, christof.paar\}@mpi-sp.org}
\IEEEauthorblockA{\IEEEauthorrefmark{3}Ruhr University Bochum, paul.staat@rub.de}
\IEEEauthorblockA{\IEEEauthorrefmark{2}Indian Institute of Technology Madras, vigneshn@smail.iitm.ac.in }}

\maketitle

\begin{abstract}
Wireless communication networks are expanding from mere information transfer to sensing: in the Integrated Sensing and Communication paradigm, existing infrastructure and signals are reused to capture physical information about the surroundings of devices, envisioned to inform high-stakes applications such as healthcare and autonomous vehicles. From a security perspective, wireless sensing inherently trusts physical observations of the radio environment. However, as in wireless communication, the open nature of radio propagation allows attackers to tamper with victim signals. In particular, attackers can use programmable metasurfaces to manipulate wireless channels and thereby mislead sensing systems. Yet, prior metasurface-based manipulations target only individual channel features such as variance or Doppler shifts. Thus, they require knowledge of the victim sensing system and produce implausible channels that are easy to detect.

In this work, we show that metasurfaces enable physically plausible manipulation of wireless sensing that is independent of the features used by the victim sensing system. Our attack reproduces genuine pre-recorded events at the physical layer and thereby synthesizes physical deepfakes of wireless channels. To this end, we develop an attack framework that controls the amplitude and phase of victim channels across time, frequency, and antennas. The key challenge here is to build a deep learning model of the interaction between metasurface and victim channel. We demonstrate that attackers can solve this challenge in several settings, e.g., by exploiting Wi-Fi Beamforming Feedback Information or replicating the victim environment. We validate the attack both in a controlled lab study on MIMO channels and with off-the-shelf Wi-Fi sensing. Our results fundamentally challenge the trustworthiness of wireless sensing.
\end{abstract}
\acresetall

\IEEEpeerreviewmaketitle

\begin{acronym}[JSONP]
\setlength{\itemsep}{0.2em}

\acro{BFI}{Beamforming Feedback Information}
\acro{CSI}{Channel State Information}
\acro{DoF}{degrees of freedom}
\acro{GPU}{Graphics Processing Unit}
\acro{ISAC}{Integrated Sensing and Communication}
\acro{LOS}[LoS]{Line-of-Sight}
\acro{MIMO}{Multiple-Input Multiple-Output}
\acro{NMSE}{Normalized Mean Squared Error}
\acro{PCB}{Printed Circuit Board}
\acro{RF}{Radio Frequency}
\acro{RIS}{Reconfigurable Intelligent Surface}
\acro{UWB}{Ultra Wideband}
\acro{VNA}{Vector Network Analyzer}
\end{acronym}

\section{Introduction}
Ubiquitous wireless networks such as cellular and \mbox{Wi-Fi} connect billions of devices in homes, offices, industrial environments, and public spaces. So far, these networks have only transmitted digital data. However, in the fundamental \ac{ISAC} paradigm shift, already deployed wireless communication infrastructure and signals will be reused for \textit{wireless sensing} to capture information about their physical surroundings. \ac{ISAC} is actively studied by standardization bodies~\cite{etsiETSIPublishesFirst2025,3gpp22837,ITU-R-M.2516-0} and equipment manufacturers~\cite{qualcommWhenWirelessSensing2024,huaweiIntegratedSensingCommunication2022}, and first field tests have been reported in China~\cite{mengNeedleHaystackTracking2026}, the EU~\cite{ericssonEricssonErillisverkotDemonstrate2026}, and the United States~\cite{attAttEricssonDemonstrate2026}.

In \mbox{Wi-Fi}, IEEE~802.11bf already standardizes sensing as a native capability~\cite{80211bf}. Technically, it builds on the \ac{CSI} that receivers estimate for every received packet, which captures how radio waves interact with the physical environment. Already today, \mbox{Wi-Fi}-based sensing can infer the activities~\cite{maWiFiSensingChannel2019}, location~\cite{zhuTuAlexaWhen2020}, vital signs~\cite{liuTrackingVitalSigns2015}, and identity~\cite{todtBFIdIdentityInference2025} of individuals. In the future, wireless sensing is envisioned to inform safety- and security-critical applications, such as collision avoidance for autonomous drones and factory robots~\cite{dongSensingService6G2023, 3gpp22837}.

From a security perspective, wireless sensing builds on the assumption that physical observations of the radio environment represent physical reality. If this assumption is violated, falsified vital signs may lead to wrong medical treatment, and spoofed motion may trigger false alarms in intrusion detection. 
This is a realistic concern, as wireless sensing faces the same security challenges as wireless communication: radio propagation is an open medium and can thus be controlled by an adversary. 
Therefore, wireless sensing is vulnerable to manipulation. 
In particular, prior work has shown that metasurfaces -- digitally configurable radio reflectors -- allow adversaries to manipulate the wireless channel in real time and mislead sensing~\cite{staatIRShieldCountermeasureAdversarial2022, jiangRISirenWirelessSensing2024, MetAegis, MetaHeart}. For example, IRShield randomly toggles metasurface configurations to obfuscate physical events~\cite{staatIRShieldCountermeasureAdversarial2022}, and RISiren switches configurations to synthesize Doppler shifts that spoof sensing events~\cite{jiangRISirenWirelessSensing2024}. Both rely on fast toggling between configurations, which produces abrupt channel discontinuities absent from natural propagation.

More fundamentally, existing spoofing attacks target individual features that sensing systems extract from the channel, such as the Doppler shift. This has two consequences. 
First, such manipulations assume a white-box setting of the victim sensing system, since manipulating the right feature requires knowing which feature the target system relies on.
Second, altering one feature in isolation leaves the remaining channel characteristics inconsistent with the spoofed event, so the synthesized channel does not correspond to any physically plausible occurrence of the sensed event.

In this work, we show that human motion and vital signs can be spoofed realistically and independently of the victim sensing system, thereby fundamentally challenging the trustworthiness of wireless sensing. We call the induced manipulations \emph{physical deepfakes} and present a novel attack framework that synthesizes them. Instead of manipulating individual features, we use a metasurface to reproduce the channel of pre-recorded genuine events, so that all features a sensing system may extract are realistic and consistent by construction. This raises two challenges. First, the attacker must find metasurface configurations that yield a target channel, for which exhaustive search is computationally intractable.
We formalize this problem and solve it by training a deep learning model of the metasurface's effect on the channel and searching it for configurations via gradient descent.
Second, this model must capture how the metasurface affects the victim channel, yet the attacker has no access to the victim sensor itself. We therefore consider five attacker settings that differ in how this channel knowledge is obtained, ranging from direct observation of the victim channel to no channel access at all.

In a lab study on \SI{160}{MHz} \ac{MIMO} channels, our framework reproduces the channels of genuine physical events across time, frequency, and antennas. 
We further find that attack difficulty is governed by the victim's measurement noise and the number of jointly synthesized channel coefficients. We then demonstrate the attack on off-the-shelf Wi-Fi devices. Here, a passive attacker exploits that \ac{BFI} is transmitted in plaintext to spoof three published \ac{BFI}-based respiration estimators, showing that this class of sensing systems is insecure. Beyond such direct observation, we study attackers with constrained channel knowledge, who instead calibrate the metasurface on a physical replica of the victim setup or via an eavesdropping antenna. A classifier trained only on genuine recordings still assigns the resulting deepfakes to the intended activity in \SI{58}{\percent} of cases with the eavesdropping antenna (chance: \SI{20}{\percent}). Finally, we show that a metasurface clones the reflection of physical objects without any knowledge of the sensing system. In summary, we make the following contributions:

\begin{itemize}

    \item We introduce physical deepfakes against wireless sensing, \ie, channel measurements that closely track physical conditions from another scene. Because they reproduce the channel itself, physical deepfakes are independent of the victim's sensing processing. We further show that prior metasurface-based attacks are easily detected in raw channel data.

	\item We introduce adversarial arbitrary channel synthesis with metasurfaces to spoof wireless sensing, which we formalize as finding metasurface configurations and solve with deep learning. We show that a metasurface attacker accurately recreates amplitude and phase channel profiles of genuine events in temporal, spectral, and spatial domains.

    \item We show that the attacker is able to calibrate the effect of the metasurface even without observing the victim channel directly.

	\item We show that \ac{BFI}-based sensing is insecure against physical manipulation as it discloses the sensing data to the attacker. We demonstrate the attack on off-the-shelf \mbox{Wi-Fi} devices.
    	
\end{itemize}

\section{Preliminaries}

\subsection{Wireless Sensing}

When a radio signal travels from a transmitter to a receiver, it propagates along multiple paths through the environment: the direct path and various reflections off walls, objects, and people.
The superposition of these paths forms the \textit{channel response}, which encodes physical changes in the propagation environment such as a person moving or breathing. \textit{Wireless sensing} exploits this by treating the channel response as a measurement of the physical surroundings. This is the principle behind radar~\cite{skolnikIntroductionRadarSystems2001}, and it can equally be realized with conventional communication signals such as \mbox{Wi-Fi}~\cite{maWiFiSensingChannel2019} or \ac{UWB}~\cite{huangUWBThroughWallImaging2010}. Communication receivers already estimate the channel, \eg, \ac{CSI} in \mbox{Wi-Fi}, for every received packet to compensate for propagation effects, and commodity chipsets expose these estimates to software~\cite{halperinToolReleaseGathering2011, xiePrecisePowerDelay2015, gringoliFreeYourCSI2019}. \mbox{Wi-Fi} devices also exchange channel estimates explicitly for beamforming. A receiver measures the channel and reports it back to the transmitter as so-called \ac{BFI}. As \ac{BFI} is transmitted unencrypted~\cite{ieee_80211_phy_standard}, any device within range can overhear the feedback and use it for sensing~\cite{beamsense}.

Building on such channel estimates, wireless sensing extracts information about physical environments and individuals, such as their location~\cite{zhuTuAlexaWhen2020}, vital signs~\cite{liuTrackingVitalSigns2015}, identity~\cite{todtBFIdIdentityInference2025}, and activities~\cite{yousefiSurveyBehaviorRecognition2017}. This has enabled applications such as fall detection in elderly care~\cite{wangWifallDevicefreeFall2016}, detection of children left alone in cars~\cite{xuWirelessAISmart2020}, and physical intrusion detection~\cite{lvRobustWLANbasedIndoor2017}.

Wireless sensing is currently moving from research into standards and products. Under the umbrella of \ac{ISAC}, physical sensing becomes a native network function built in from the ground up rather than an opportunistic reuse of communication signals~\cite{liuIntegratedSensingCommunications2022, chowdhury6GWirelessCommunication2020}. Upcoming 6G~\cite{dongSensingService6G2023} and \mbox{Wi-Fi}~\cite{80211bf} standards will enable base stations, routers, and smartphones to analyze their physical surroundings, and the International Telecommunication Union acknowledges sensing as a key technology trend beyond 2030~\cite{ITU-R-M.2516-0}. 3GPP, the standardization body behind cellular standards worldwide, has already studied 32 sensing use cases, including safety-critical applications such as collision avoidance for autonomous drones and factory robots~\cite{3gpp22837}. First field tests are already running in the EU~\cite{ericssonEricssonErillisverkotDemonstrate2026}, China~\cite{mengNeedleHaystackTracking2026}, and the United States~\cite{attAttEricssonDemonstrate2026}, and home security products based on \mbox{Wi-Fi} sensing are commercially available~\cite{comcastLaunchesXfinity2026}.

\subsection{Metasurface Technology}
\label{sec:prelim_metasurface}

Around 2014, programmable metasurfaces emerged from metamaterial physics to tailor radio propagation environments~\cite{kainaShapingComplexMicrowave2014}. In wireless communication research, they are typically referred to as \acp{RIS} and are considered a potential building block for future 6G networks~\cite{tataria6GWirelessSystems2021}. Today, metasurfaces have matured into commercial products~\cite{Greenerwave2025, NovoFlectReconfigurableIntelligent2025}. By electronically controlling how they reflect impinging waves, metasurfaces make the propagation medium itself a tunable resource~\cite{jiangRoad6GComprehensive2021}, \eg, to redirect radio waves, improve coverage, or eliminate blind spots~\cite{liaskosNovelCommunicationParadigm2019, liuReconfigurableIntelligentSurfaces2021, basarWirelessCommunicationsReconfigurable2019}.

A typical metasurface is a planar array of $L$~identical unit-cell reflectors printed on a low-cost \ac{PCB}~\cite{ranaReviewPaperHardware2023}. The reflection phase of each element is tuned independently through digital logic signals, \eg, from a microcontroller. In the common case of 1-bit control, each element switches between two reflection phase states, \SI{0}{\degree} and \SI{180}{\degree}, corresponding to reflection coefficients of $+1$ and $-1$. The metasurface configuration is thus a binary vector $\mathbf{c} \in \{0,1\}^L$, and every surface configuration takes physical effect as a distinct reflection pattern.

Which configuration yields a desired behavior, such as maximizing signal power between a transmitter and a receiver, depends on the surrounding radio environment: the reflected components must add coherently with the direct propagation paths. Since a binary metasurface offers $2^L$ configurations, practical control relies on iterative optimization with feedback from channel measurements~\cite{feng2021optimization, zou2021robust, tewesIRSenabledBreathTracking2022, kainaShapingComplexMicrowave2014, peiRISaidedWirelessCommunications2021}.

\section{Physical Deepfake Attacks against Wireless Sensing}
\label{sec:model}

In this section, we introduce the system and attacker models and define the attack objective. We then introduce physical deepfakes as a distinct class of attacks to mislead wireless sensing systems, in which an adversary physically manipulates the victim environment.

\subsection{System and Attacker Model}
\label{sec:system_model}

\begin{figure}[t]
  \centering
  \includegraphics[width=0.845\columnwidth]{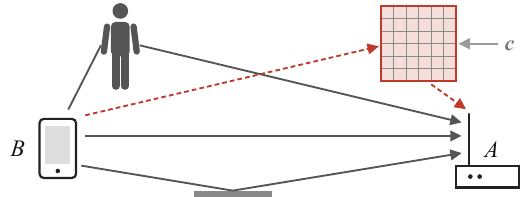}
  \caption{System model. Transmitter~$B$ reaches the victim sensor~$A$ over several propagation paths, including benign and malicious reflections.}
  \label{fig:system_model}
\end{figure}

\Paragraph{System Model}
We consider bistatic wireless sensing where two wireless devices are deployed in an indoor environment, \eg, a home or an office. The device $A$ is the victim sensor that receives sensing signals transmitted by device~$B$. $A$ estimates the wireless channel $\mathbf{H}$ from $B$'s signals and infers physical events from it. Depending on the sensing waveform and hardware capabilities, $A$ may observe the channel over time, frequency, and spatial domains. $\mathbf{H}$ is the superposition of all propagation paths from $B$ to $A$, $\mathbf{H} = \mathbf{H}_\text{static} + \mathbf{H}_\text{event}$. The direct path and wall reflections are denoted by $\mathbf{H}_\text{static}$. The propagation paths from which $A$ infers physical events, such as reflections off a moving person, form the component $\mathbf{H}_\text{event}$.

\Paragraph{Attacker Model}
The attacker has physical access to the victim environment and can place a metasurface that affects the channel $\mathbf{H}$ between $A$ and $B$. The attacker can inspect the victim environment and thus knows the sensing hardware and the geometry of the victim setup, including the positions of $A$ and $B$. Further, the attacker can eavesdrop on all wireless messages transmitted in the victim environment to observe physical-layer signals but cannot read encrypted data. The attacker has no access to $A$: they can neither read nor modify $A$'s channel estimates and they do not know how $A$ infers physical events, \eg, the sensing processing, features, and inference function.

\Paragraph{Metasurface}
The attacker's metasurface consists of $L$ elements and adds a third component to the victim channel,
\begin{equation}
  \mathbf{H} = \mathbf{H}_\text{static} + \mathbf{H}_\text{event} + \mathbf{H}_\text{meta}(\mathbf{c}),
  \label{eq:channel_components}
\end{equation}
where $\mathbf{c} \in \{0,1\}^L$ is the metasurface configuration. $\mathbf{H}_\text{meta}$ is the superposition of $L$ concatenated channels from $B$ to the metasurface elements and on to $A$,
\begin{equation}
    \mathbf{H}_\text{meta}(\mathbf{c}) = \sum_{l=1}^L r(c_l)\, \mathbf{H}_l,
  \label{eq:h_meta}
\end{equation}
where $\mathbf{H}_l$ is the channel from $B$ via element~$l$ to $A$ and $r(c_l)$ is the reflection coefficient of element~$l$ in state $c_l$, \ie, $r(0) = +1$ and $r(1) = -1$, \cf~\autoref{sec:prelim_metasurface}.
The attacker switches $\mathbf{c}$ over time at least as fast as $A$ samples the channel to partially control how $A$ observes $\mathbf{H}$. The attacker cannot remove the static paths, \ie, they shape the channel relative to $\mathbf{H}_\text{static}$.

\Paragraph{Attacker Goal}
By reconfiguring the metasurface, the attacker aims to make $A$ infer a target event that does not physically occur, \eg, a person walking through an empty room.

\Paragraph{Attacker Settings}
To synthesize a target channel, the attacker needs to know how a configuration $\mathbf{c}$ affects the channel at $A$. Since the attacker has no access to $A$, they cannot directly observe this effect. We therefore distinguish five attacker settings that refine the attacker model above and differ in how the attacker obtains this channel knowledge, as summarized in \autoref{tab:attacker_settings}. The settings are ordered by decreasing channel knowledge, and our evaluation follows this order.

\begin{table*}[t]
\centering
\caption{Attacker settings and respective capabilities to obtain channel knowledge.
}
\label{tab:attacker_settings}
\renewcommand{\arraystretch}{1.2}
\small
\begin{tabularx}{\textwidth}{@{}l l >{\hsize=0.7\hsize}X >{\hsize=1.3\hsize}X c@{}}
\toprule
\textbf{Family} & \textbf{Setting} & \textbf{Attacker control} & \textbf{Source of channel knowledge} & \textbf{Sec.} \\
\midrule
\multirow{2}{*}{\shortstack[l]{Observation}}
& \textbf{S1} Dishonest partner & Controls $B$ & $B$ can directly measure the reciprocal channel & \ref{sec:feasibility} \\
& \textbf{S2} \ac{BFI} eavesdropper & Passive eavesdropping & Direct access through unencrypted \ac{BFI} & \ref{sec:bfi} \\
\midrule
\multirow{2}{*}{\shortstack[l]{Estimation}}
& \textbf{S3} Physical replica & Rebuilds setup offline & Measured on an offline replica of the victim setup & \ref{sec:constrained:s3} \\
& \textbf{S4} Symmetric proxy & Eavesdropping antenna $E$ & Inferred via geometric symmetry between $A$ and $E$& \ref{sec:constrained:s4} \\
\midrule
Cloning
& \textbf{S5} Environment cloning & Initial environment access & None (metasurface mimics a legitimate environment) & \ref{sec:cloning} \\
\bottomrule
\end{tabularx}
\end{table*}

\subsection{Physical Deepfakes}
\label{sec:formalization}

We now define physical deepfakes against wireless sensing by abstracting a general sensing model, which we then instantiate with metasurface-based wireless sensing manipulation.

\subsubsection{General Sensing Model}
We abstract the setting from \autoref{sec:system_model} into a chain of
mappings from a physical event $X$ to an inferred event $\hat{X}$. While the model is designed with wireless sensing in mind, it applies equally to arbitrary sensing systems, such as cameras, lidar, or radar, and therefore allows us to define the attack independently of the sensing modality:
\begin{equation}
  X \xrightarrow{\;\pi\;} Y
    \xrightarrow{\;\varepsilon\;} \mathbf{M}
    \xrightarrow{\;\Phi\;} f
    \xrightarrow{\;g\;} \hat{X},
  \label{eq:pipeline}
\end{equation}
where the physical transformation $\pi$ maps an event $X$, \eg, a person walking, to the signal $Y$ at the sensor, \eg, by radio propagation from $B$ to $A$. The measurement function $\varepsilon$ turns $Y$ into the measurement $\mathbf{M}$, \eg, $A$'s channel estimate. The processing $\Phi$ extracts features $f$, \eg, Doppler shifts, from which the inference function $g$, \eg, an activity classifier, infers the event $\hat{X}$.

The measurement is a sequence of $T$~snapshots
$\mathbf{M} = (\mathbf{M}_1, \dots, \mathbf{M}_T)$, \eg, channel estimates
from successive packets, where each $\mathbf{M}_i \in \mathbb{C}^{d}$ captures the $d$ \ac{DoF} of the sensor.

\subsubsection{Attack Objective}
Let $X$ denote the physical event that actually occurs and
$\tilde{X} \neq X$ the target event the attacker aims to make the system
infer. We write $\mathbf{M}(X) = \varepsilon(\pi(X))$ for the measurement of an event~$X$. The attacker is unable to present $\tilde{X}$ physically. Instead, they manipulate the environment with a configuration $\mathbf{c}$, \eg, the metasurface configuration in \autoref{sec:system_model}, and thereby synthesize the measurement $\mathbf{M}(\mathbf{c}) = \varepsilon(\pi(X; \mathbf{c}))$. We assume that the environment is static during the attack, \ie, no genuine event occurs, so that the synthesized measurement depends only on $\mathbf{c}$. We distinguish three levels of attack success, in increasing order of strength:
\begin{align}
  \text{(inference level)}\quad
    & g(\Phi(\mathbf{M}(\mathbf{c}))) = g(\Phi(\mathbf{M}(\tilde{X}))), \\
  \text{(feature level)}\quad
    & \Phi(\mathbf{M}(\mathbf{c})) \approx \Phi(\mathbf{M}(\tilde{X})), \\
  \text{(measurement level)}\quad
    & D\big(\mathbf{M}(\mathbf{c}), \mathbf{M}(\tilde{X})\big)
      \leq \delta_\varepsilon,
\end{align}
where $D(\cdot,\cdot)$ is a distance metric and
$\delta_\varepsilon$ a tolerance determined by the quality of the measurement function.

\takeaway[bottom=2pt]{\setlength{\topsep}{0pt}\setlength{\partopsep}{0pt}\begin{definition}[Physical deepfake]
\label{def:physical_deepfake}
A \emph{physical deepfake} is a measurement $\mathbf{M}(\mathbf{c})$ that an attacker induces by physically manipulating the environment aiming to match the measurement of a target event~$\tilde{X}$. It succeeds at the measurement level if $D\big(\mathbf{M}(\mathbf{c}), \mathbf{M}(\tilde{X})\big) \leq \delta_\varepsilon$.
\end{definition}}

We adopt the term from recent work on humanoid robots that imitate human motion~\cite{kim2026irobot} and refer to the attempt to produce a physical deepfake as a \emph{physical deepfake attack}. Like media deepfakes, a physical deepfake is synthesized with deep learning and impersonates a genuine event. By construction, it is independent of $\Phi$ and $g$ and cannot be detected by any downstream mechanism that operates solely on $\mathbf{M}$. It thus constitutes the attack goal and matches our attacker model, in which the attacker does not possess knowledge of $\Phi$ and $g$. Unlike physical adversarial examples~\cite{kurakinAdversarialExamplesPhysical2016, eykholtRobustPhysicalWorldAttacks2018}, which keep the measurement close to that of the actual event and exploit weaknesses of $g$, a physical deepfake reproduces the measurement of the target event and thus deceives even an ideal inference function.

\subsubsection{Realizing Physical Deepfakes with Metasurfaces}
In the setting from \autoref{sec:system_model}, the measurement $\mathbf{M}$ is $A$'s channel estimate $\mathbf{H} = (\mathbf{H}_1, \dots, \mathbf{H}_T)$ with $\mathbf{H}_i \in \mathbb{C}^{s \times k}$ for $s$ spatial links and $k$ subcarriers, \ie, $d = s \cdot k$. The metasurface configuration $\mathbf{c}$ acts on the propagation $\pi$ by adding $\mathbf{H}_\text{meta}(\mathbf{c})$ to the channel in Eq.~\eqref{eq:channel_components}. For simplicity, we assume a static environment during the attack, $\mathbf{H}_\text{event} = \mathbf{0}$, and a physical deepfake requires
\begin{equation}
  \mathbf{H}_\text{static} + \mathbf{H}_\text{meta}(\mathbf{c}) \approx \mathbf{H}_\text{static} + \mathbf{H}_\text{event}(\tilde{X}),
  \label{eq:deepfake_channel}
\end{equation}
\ie, the metasurface paths stand in for the paths a genuine occurrence of the target event~$\tilde{X}$ would induce.

To synthesize a target channel, the attacker needs a model $\hat{\mathbf{H}}(\mathbf{c})$ that predicts the channel at $A$ for a given metasurface configuration $\mathbf{c}$. We consider several attacker settings for obtaining this model, as summarized in \autoref{tab:attacker_settings}.

\section{Synthesizing Physical Deepfakes with Metasurfaces}
\label{sec:framework}

\begin{figure*}[t]
  \centering
  \includegraphics[width=\textwidth]{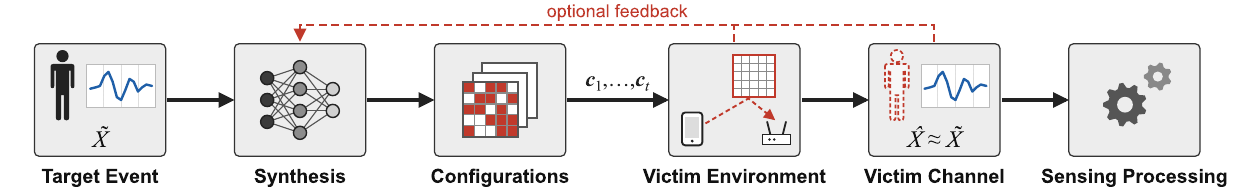}
  \caption{Physical deepfake attack flow. From a recording of the target event $\tilde{X}$, the attacker synthesizes a sequence of metasurface configurations $\mathbf{c}_1,\dots,\mathbf{c}_T$. Replayed in the victim environment, they shape the victim channel such that the sensing processing infers $\hat{X} \approx \tilde{X}$. Feedback from the victim environment is optional.}
  \label{fig:attack_flow}
\end{figure*}

A physical deepfake requires the attacker to make $A$'s channel follow the target channel $\mathbf{H}_\text{static} + \mathbf{H}_\text{event}(\tilde{X})$. The attacker thus has to find a sequence of metasurface configurations $(\mathbf{c}_i)_{i=1}^T$ that reproduces a recording of the target channel. With $2^L$ possible configurations, exhaustive search is intractable. Moreover, the effect of each configuration on $A$'s channel depends on the victim environment, which the attacker cannot observe through $A$. Finally, the sequence must reproduce the target jointly across time, frequency, and antennas. In the following, we discuss how the attacker can resolve these challenges.

\subsection{Forward Surrogate}

To search for configurations, the attacker needs to predict the channel at $A$ for a given configuration.
The idealized physical model in Eq.~\eqref{eq:h_meta} requires the per-element channels $\mathbf{H}_l$, which the attacker cannot measure, and neglects the coupling between adjacent elements. We therefore learn this mapping from channel measurements with a neural network, the \emph{forward surrogate} $\hat{\mathbf{H}}(\mathbf{c})$.

The surrogate maps a configuration $\mathbf{c}$ to the channel $\hat{\mathbf{H}}(\mathbf{c}) \in \mathbb{C}^{s \times k}$ at $A$. It sums two parallel branches: a linear branch that mirrors Eq.~\eqref{eq:h_meta}, and a convolutional branch that treats $\mathbf{c}$ as a $16\times16$ aperture image, as per \autoref{sec:setup}, and captures the coupling between adjacent elements. We train the surrogate on channel measurements of known configurations.

We consider different attacker settings, depending on how much channel knowledge the attacker can observe for each configuration, as summarized in \autoref{tab:attacker_settings}. When the victim channel is observable, the attacker trains the surrogate on the victim channel itself, obtained either through $B$ (S1) or eavesdropped \ac{BFI} (S2). When the victim channel is out of reach, the attacker trains on another wireless channel (either a channel replica or a helper channel) and then estimates the victim environment with this surrogate (S3, S4). With only one-time initial access to the victim environment and no access to the victim channel (S5), the attacker profiles the reflection of a part of the environment and clones it with a metasurface trained on its own setup. We curate a dataset of known configurations covering random patterns, their bit-flip neighbors, sparse and structured patterns, and beam-steering configurations, so that the impact of each element of the metasurface is observed in the dataset. 
If only a few measurements are possible, we fine-tune a surrogate pretrained in another environment following~\cite{kumar2022finetuning} (see \autoref{sec:surr_impact}). 

We consider a surrogate accurate if its average distance $D\big(\mathbf{H}(\mathbf{c}), \hat{\mathbf{H}}(\mathbf{c})\big)$ on held-out configurations stays within the tolerance $\delta_\varepsilon$ given by the system noise. 
We will further discuss the required training effort in \autoref{sec:surr_impact} and the model architecture and training details are in Appendix~\ref{app:surrogate}.

\subsection{Configuration Search}
\label{sec:config_search}
We obtain a configuration $\mathbf{c}$ for a target channel $\mathbf{H} \in \mathbb{C}^{s \times k}$ with an iterative gradient-based search on the forward surrogate, where we utilize the gradient of the distance metric $D\big(\mathbf{H}, \hat{\mathbf{H}}(\mathbf{c})\big)$, which is used as a loss function.
Because the surrogate is differentiable and runs on a \ac{GPU}, this search strategy is cheap and can be parallelized. 
Another strength of the approach is that the search is capable of generating metasurface configurations that are not in the training dataset.
We treat the binary states of $\mathbf{c}$ as trainable parameters and freeze the surrogate weights. We optimize one continuous logit per element and threshold the logits to a binary configuration before each surrogate evaluation, so that every evaluated configuration is realizable. Gradients pass through the threshold with a straight-through estimator. We refine the best configuration with single-element and pair flips that lower the error. Search parameters and refinement details are given in  Appendix~\ref{app:surrogate}.

As detailed in \autoref{sec:system_model}, we can design channel responses across time, frequency, and space to attack a variety of victim sensors. For a channel estimate  $\mathbf{H}_i \in \mathbb{C}^{s \times k}$, where $1 \leq i \leq T$, we may focus on a subset of spatial links $\mathcal{S} \subseteq \{1, \dots, s\}$ and a subset of subcarriers $\mathcal{K} \subseteq \{1, \dots, k\}$. The inverse design and distance computation are then performed only considering the channel estimate $\mathbf{H}_i' \in \mathbb{C}^{|\mathcal{S}| \times |\mathcal{K}|}$. This allows for the design of spoofing attacks in specific frequency bands or across spatial links. The relationship between synthesis quality and the number of \ac{DoF} for a particular frame is described in \autoref{sec:driving-factors}.
The inverse design is performed independently for each time step, producing a configuration sequence $(\mathbf{c}_i)_{i=1}^T$ that is replayed on the metasurface at the recording's original sampling rate. Unlike toggling between fixed configurations, each configuration reproduces the target channel at its time step, so the observed channel follows the natural variation of the target event instead of jumping between fixed states. Coupling frames across time would build an assumption about the target's temporal variation into the attack, which would limit the range of synthesizable activities.

\subsection{Experimental Setup}
\label{sec:setup}

We describe the hardware shared across our experiments. Setups specific to individual experiments are described in the respective sections.

\Paragraph{Metasurface}
The attacker's metasurface follows the open-source design of Heinrichs~\etal~\cite{heinrichsOpenSourceReconfigurable2023}. 
It is an FR4 \ac{PCB} with $L = 256$ elements arranged in a $16 \times 16$ grid and measures \SI{36.0}{\cm}~$\times$~\SI{24.7}{\cm}. Each element has binary phase control and is optimized for the \SI{5}{\GHz} \mbox{Wi-Fi} band. We program a configuration $\mathbf{c}$ via USB serial communication, which sets each element~$l$ to state~$c_l = 0$ or $c_l = 1$, \ie, to a reflection phase of \SI{0}{\degree} or \SI{180}{\degree} ($r(c_l) = \pm 1$ in Eq.~\eqref{eq:h_meta}). Further technical details are given in \cite{heinrichsOpenSourceReconfigurable2023}.

\begin{figure}
    \centering
    \includegraphics[width=1.0\linewidth]{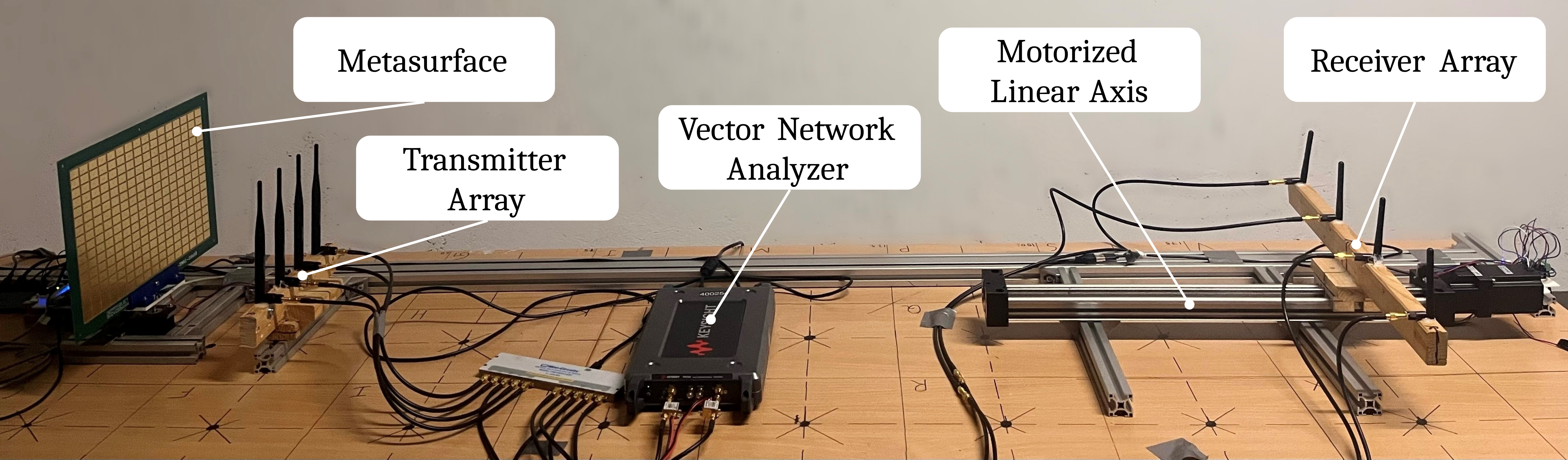}
    \caption{A \ac{VNA}-based lab setup for measuring a $4 \times 4$ \ac{MIMO} channel, including adversarial metasurface propagation paths.}
    \label{fig:setup}
\end{figure}

\Paragraph{Lab Setup}
Our lab setup (\autoref{fig:setup}) measures a $4 \times 4$ \ac{MIMO} channel, \ie, $s = 16$ spatial links. A Keysight P9372A \ac{VNA} in conjunction with \ac{RF} switches records the complex-valued frequency response of each link. For each link, we measure $129$ frequency points over a \SI{160}{MHz} band centered at \SI{5.57}{GHz}, with an intermediate-frequency bandwidth of \SI{500}{kHz}. The measurement of a single frame of the \ac{MIMO} channel takes around \SI{0.13}{s}. The metasurface is placed \SI{0.2}{m} from the transmit array and adds the attacker-controlled path to the channel. Motorized stages vary the propagation conditions. A linear axis sets the distance between the transmit and receive arrays from \SI{1}{m} to \SI{1.5}{m} in steps of \SI{10}{cm}, and a rotation stage sets the angle of the metasurface relative to the transmit array to \SI{0}{\degree}, \SI{30}{\degree}, or \SI{45}{\degree}. To characterize various environments, we measure $150{,}800$ known metasurface configurations for every combination of axial distance and metasurface angle. 

\Paragraph{\mbox{Wi-Fi} Devices}
For experiments on commodity hardware, we use ESP32-C5 devices and a Netgear Nighthawk~RAX50 router, which operate an IEEE~802.11ac network with a channel bandwidth of \SI{20}{MHz}. All networks are configured to \mbox{Wi-Fi} channel 112, matching the effective bandwidth of the metasurface. In settings S3 and S4, ESP32-C5 devices serve as both transmitter and responder and \ac{CSI} is extracted from the receiver. In setting S2, an ESP32-C5 acts as a legitimate participant and thus takes part in the beamforming protocol, where \ac{BFI} can be eavesdropped in plaintext.

\Paragraph{Metrics}
For a target frame $\mathbf{H}_i$ and a synthesized channel $\hat{\mathbf{H}}_i$, where $\mathbf{H}_i, \hat{\mathbf{H}}_i \in \mathbb{C}^{s \times k}$, we use the \ac{NMSE} as the distance metric:
\begin{equation}
\mathrm{NMSE}_i(\hat{\mathbf{H}}_i, \mathbf{H}_i) =
\frac{\bigl\lVert \hat{\mathbf{H}}_i - \mathbf{H}_i \bigr\rVert_F^2}
     {\bigl\lVert \mathbf{H}_i \bigr\rVert_F^2},
\label{eq:nmse}
\end{equation}
where $\lVert \cdot \rVert_F$ denotes the Frobenius norm. The error is thus normalized by the energy of the respective target frame, which makes the metric independent of the absolute channel gain. To summarize the result over a set of $N$ frames, we convert the per-frame values to decibels and average them:
\begin{equation}
\mathrm{NMSE} = \frac{1}{N} \sum_{i=1}^{N} 10 \log_{10} \mathrm{NMSE}_i(\hat{\mathbf{H}}_i, \mathbf{H}_i).
\label{eq:nmse_avg}
\end{equation}
Averaging in the logarithmic domain corresponds to the geometric mean of the per-frame values and is less sensitive to individual outlier frames. To put these values into context, we report a noise floor computed with the same metric, namely the \ac{NMSE} between repeated measurements of a static environment, which captures the measurement noise of the system. We then set the tolerance level $\delta_\varepsilon$ to the 99th percentile of the resulting distribution.

\section{Attack Feasibility}
\label{sec:feasibility}

\begin{figure}[t]
  \centering
  \includegraphics[width=1.0\columnwidth]{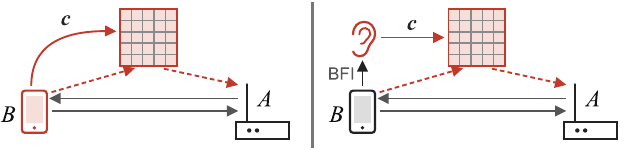}
  \caption{Attacker settings with observed channel knowledge. Left~(S1): the attacker controls $B$, which observes the reciprocal channel to $A$ and drives the metasurface configuration~$\mathbf{c}$. Right~(S2): $B$ is legitimate and emits unencrypted \ac{BFI}, which a passive attacker overhears to derive $\mathbf{c}$.}
  \label{fig:observation_settings}
\end{figure}

We first evaluate physical deepfakes in setting~S1, in which the attacker controls the transmitter~$B$ (see~\autoref{fig:observation_settings}). Because the channel is reciprocal, $B$ measures the channel to $A$ and thus observes the effect of each metasurface configuration on the victim channel. This setting arises wherever the victim senses with signals of devices it does not control, \eg, a router in a public \mbox{Wi-Fi} network that senses with the packets of unknown client devices. Any device that joins the network can then act as~$B$. Our \ac{VNA}-based lab setup mimics this setting.

\subsection{Arbitrary Channel Synthesis with Physical Deepfakes}

\begin{figure}[t]
\centering
\begin{subfigure}{\columnwidth}
    \includegraphics[width=\linewidth]{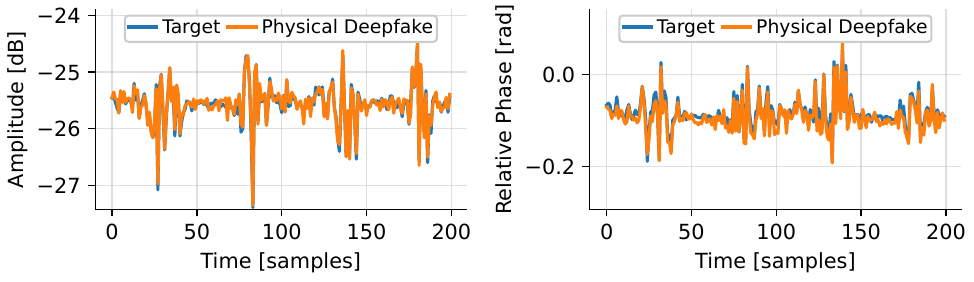}
    \caption{Walking}
    \label{fig:walking}
\end{subfigure}
\begin{subfigure}{\columnwidth}
    \includegraphics[width=\linewidth]{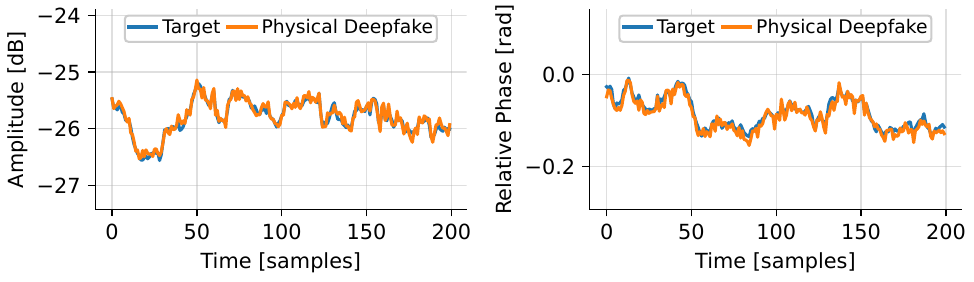}
    \caption{Clapping}
    \label{fig:clapping}
\end{subfigure}
\begin{subfigure}{\columnwidth}
    \includegraphics[width=\linewidth]{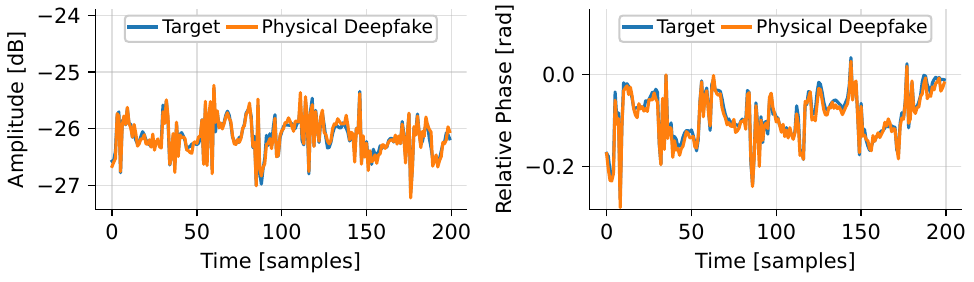}
    \caption{Sitting down}
    \label{fig:sitting}
\end{subfigure}
\caption{Examples of time-domain physical deepfakes, synthesizing both channel amplitude and phase profiles for several target sensing events.}
\label{fig:actions}
\end{figure}

In our first experiment, we use the lab sensing setup to record three target events (an individual walking, clapping, and sitting down), each repeated multiple times. We then place the adversarial metasurface in the environment, train the forward surrogate on channel measurements of known configurations, and synthesize a configuration sequence for each recording. We replay this sequence on the metasurface and compare the resulting channel against the target. In this experiment, each configuration synthesizes a single \ac{DoF}, \ie, the complex channel coefficient of one spatial link at one frequency for one time sample. \autoref{fig:actions} shows the resulting amplitude and phase over time. The replayed deepfake channels closely follow the original recording. The \ac{NMSE} of the deepfake channels lies between \SI{-37.3}{dB} and \SI{-36.3}{dB}, which is close to \SI{-41.1}{dB}, the repeatability limit of our setup.

\begin{figure}[t]
\centering
\begin{subfigure}{\columnwidth}
    \includegraphics[width=\linewidth]{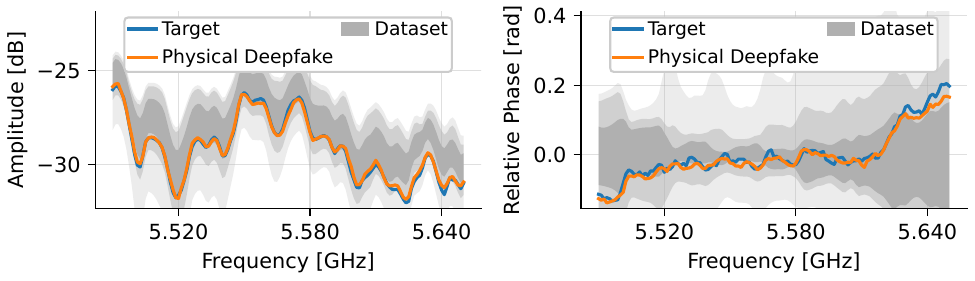}
    \caption{Multiple frequency bands}
    \label{fig:sensing_freq}
\end{subfigure}
\\
\begin{subfigure}{\columnwidth}
    \includegraphics[width=\linewidth]{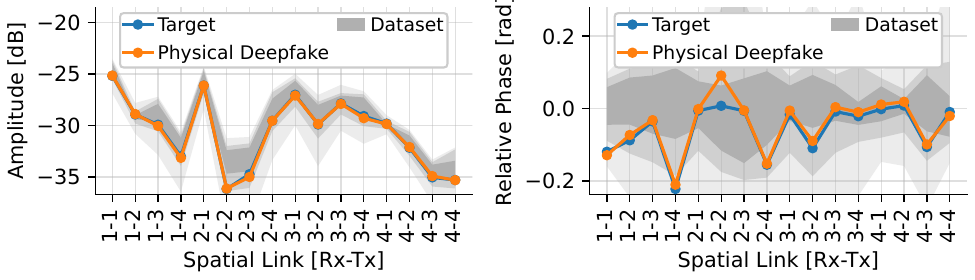}
    \caption{Multiple spatial links}
    \label{fig:sensing_space}
\end{subfigure}
\caption{Examples of frequency-domain and spatial-domain physical deepfakes of a snapshot of the Clapping event.}
\label{fig:sensing_domains}
\end{figure}

Sensing systems observe the channel across frequency and space, which makes the metasurface synthesis harder: in the time domain, each sample has its own configuration, whereas across frequency and space, a single configuration must match multiple \ac{DoF} at once. \autoref{fig:sensing_domains} shows the synthesis for one time-sample of the Clapping event. Here, the synthesis reaches an \ac{NMSE} of \SI{-34.1}{dB} across the full \SI{160}{MHz} band on a single spatial link, and \SI{-35.2}{dB} across all 16 spatial links for a single frequency.

\begin{figure}[t]
\centering
\begin{subfigure}{0.32\columnwidth}
    \includegraphics[width=\linewidth]{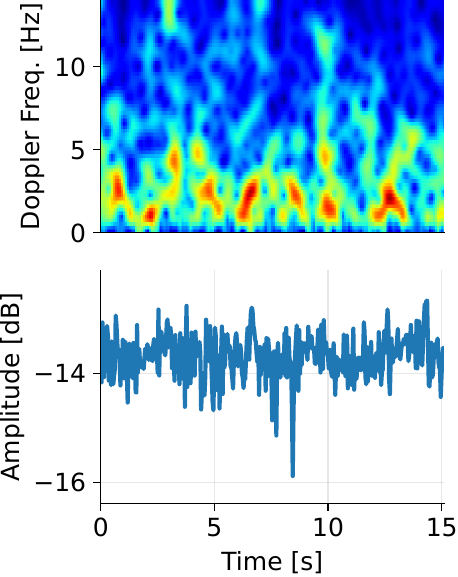}
    \caption{Target}
    \label{fig:risiren_target}
\end{subfigure}\hfill
\begin{subfigure}{0.32\columnwidth}
    \includegraphics[width=\linewidth]{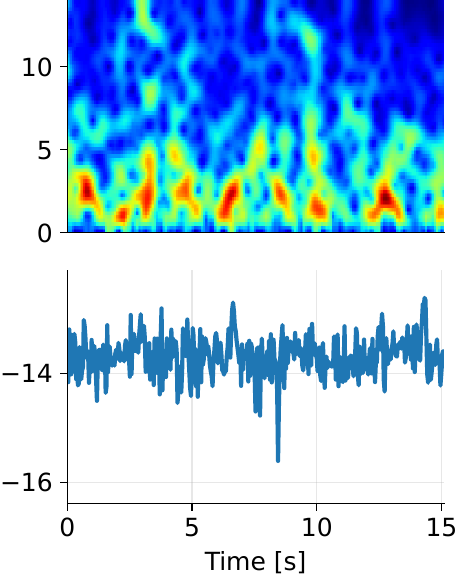}
    \caption{Phy. Deepfake}
    \label{fig:risiren_deepfake}
\end{subfigure}\hfill
\begin{subfigure}{0.32\columnwidth}
    \includegraphics[width=\linewidth]{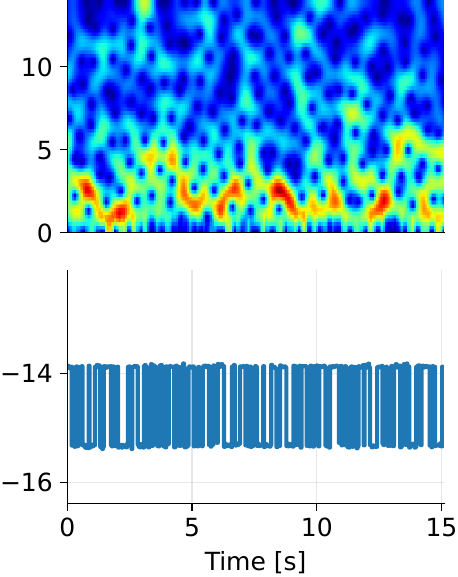}
    \caption{RISiren \cite{jiangRISirenWirelessSensing2024}}
    \label{fig:risiren}
\end{subfigure}

\caption{Comparison of our attack~(b) with RISiren feature manipulation~(c). While RISiren spoofs the Doppler frequency spectrum, the physical deepfake tracks the raw channel measurement.}
\label{fig:risiren_comparison}
\end{figure}

Prior metasurface attacks manipulate the features a sensing system extracts rather than the raw channel, which exposes them to a defender who inspects the raw measurement. For example, RISiren~\cite{jiangRISirenWirelessSensing2024} spoofs the Doppler spectrum by toggling between two metasurface configurations. \autoref{fig:risiren_comparison} compares RISiren with a physical deepfake of the same target event. In the Doppler domain, the toggling reproduces the intended signature, so RISiren would mislead a Doppler-based classifier. In the raw channel, however, the toggling leaves a rectangular pattern that does not occur in natural propagation. In contrast, the physical deepfake follows the target in both domains.

\takeaway{Metasurfaces reproduce the raw channel of genuine events across time, frequency, and space. Unlike feature-level manipulation, a physical deepfake matches the target in the raw channel and thus in every feature derived from it.}

\subsection{Surrogate Training Effort}
\label{sec:surr_impact}
\begin{figure}[t]
\centering
\begin{subfigure}{0.47\columnwidth}
    \includegraphics[width=\linewidth]{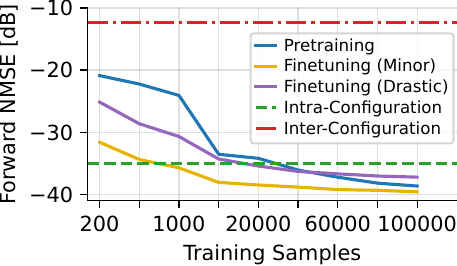}
    \caption{}
    \label{fig:forward_nsamples}
\end{subfigure}\hfill
\begin{subfigure}{0.5\columnwidth}
    \includegraphics[width=\linewidth]{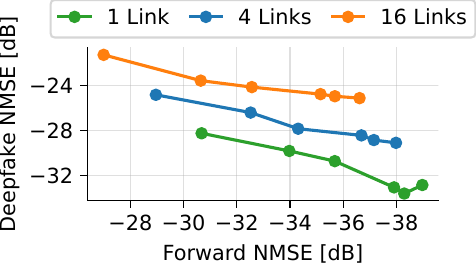}
    \caption{}
    \label{fig:forward_backward}
\end{subfigure}
\caption{Performance evaluation of the deep learning surrogate model. (a) The error of the surrogate model predicting the channel depending on the amount of training samples used. (b) Relationship between forward error and synthesis quality with different numbers of \ac{DoF} synthesized, for the Clapping activity.}
\label{fig:surrogate_perf}
\end{figure}

A key step of the attack is to fit the forward surrogate to the victim environment. \autoref{fig:forward_nsamples} reports this effort as the number of training samples needed to reach a given prediction error. The upper bound is the inter-configuration spread, \ie, the average \ac{NMSE} between the channels of two different configurations, at which a surrogate predicts no better than the channel of an arbitrary other configuration. The lower bound is the intra-configuration spread, \ie, the average \ac{NMSE} between repeated measurements of the same configuration. This spread reflects the system noise and marks the error an attacker aims for.

Trained from scratch, the surrogate needs $30{,}000$ samples to reach the noise floor, with the steepest decrease between $1{,}000$ and $10{,}000$ samples. This method scales well to new environments because fine-tuning reduces the training burden. After a minor physical change in which we displace the receive antennas along the linear axis, a surrogate pretrained on $135{,}720$ samples reaches the lower bound with fewer than $1{,}000$ samples. After a drastic physical change, in which we displace the transmit antennas and shift and rotate the receive array, the fine-tuned surrogate still needs fewer samples than a surrogate trained from scratch.

Because the configuration search operates on the surrogate, deepfake quality follows surrogate accuracy: \autoref{fig:forward_backward} shows that a surrogate with a lower prediction error yields physical deepfakes with a lower \ac{NMSE}. For a given surrogate, the deepfake error grows with the number of jointly synthesized \ac{DoF}. This growth reflects the error of the configuration search, which must match more channel coefficients with a single configuration.

\subsection{Attack Success and Synthesis Difficulty}
\label{sec:driving-factors}

\begin{figure}[t]
\centering
\begin{subfigure}{0.333\columnwidth}
    \includegraphics[width=\linewidth]{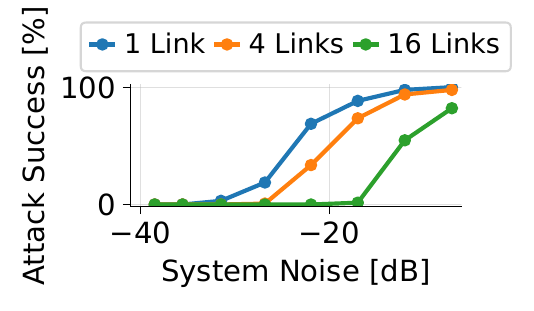}
    \caption{\SI{20}{MHz}}
    \label{fig:noise1}
\end{subfigure}\hfill
\begin{subfigure}{0.333\columnwidth}
    \includegraphics[width=\linewidth]{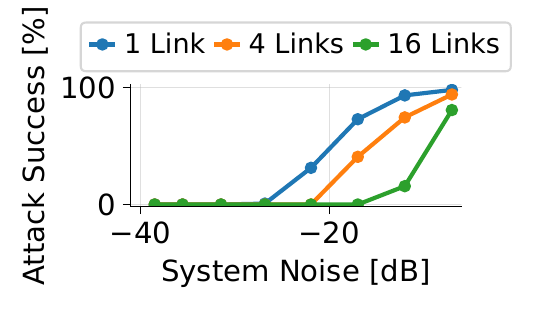}
    \caption{\SI{80}{MHz}}
    \label{fig:noise2}
\end{subfigure}\hfill
\begin{subfigure}{0.333\columnwidth}
    \includegraphics[width=\linewidth]{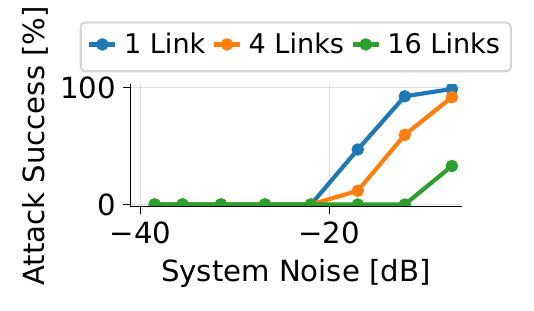}
    \caption{\SI{160}{MHz}}
    \label{fig:noise3}
\end{subfigure}
\caption{Attack success rate versus the victim's system noise for 1, 4, and 16 spatial links at three bandwidths.}
\label{fig:noiseSuccess}
\end{figure}

We evaluate attack success against the victim's measurement quality, which we quantify as its system noise, \ie, the \ac{NMSE} between repeated measurements of the same channel. To emulate noisier victim sensing, we add complex Gaussian noise to all measurements and raise the system noise from the repeatability limit up to \SI{-7}{dB}. The noise affects the attacker's training data and target as well as the synthesized channel that the victim observes.

A physical deepfake succeeds if the victim cannot distinguish it from a repeated measurement of the genuine channel, \ie, if its \ac{NMSE} to the target stays below the tolerance $\delta_\varepsilon$. For each noise level, we set $\delta_\varepsilon$ to the 99th percentile of the system noise, which corresponds to a detector with a false positive rate of \SI{1}{\percent}. \autoref{fig:noiseSuccess} reports the fraction of target samples for which the attack succeeds, for bandwidths of \SI{20}{MHz}, \SI{80}{MHz}, and \SI{160}{MHz} and for 1, 4, and 16 spatial links. Synthesis and evaluation are restricted to the respective bandwidth and spatial links. We find that the victim's measurement quality is a critical factor for attack success as the success rate increases with the system noise. A low-noise victim tolerates almost no deviation from the target. With more noise, the victim's own measurements become less repeatable, the tolerance $\delta_\varepsilon$ increases, and the attack success rate approaches \SI{100}{\percent}. Attack success further decreases with the number of jointly synthesized \ac{DoF}. More spatial links and a wider band both shift the success curve toward higher noise.

\takeaway{Attack difficulty grows with the victim's measurement quality and with the number of jointly synthesized \ac{DoF}.}

\begin{figure}
    \centering
    \includegraphics[width=\linewidth]{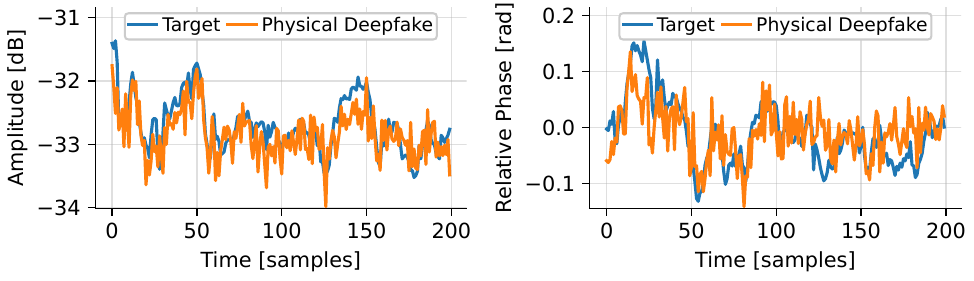}
    \caption{Time-domain physical deepfake of Clapping across all 16 spatial links at \SI{160}{MHz}, with an \ac{NMSE} of \SI{-26.7}{dB}. It misses the success criterion but reproduces the coarse structure of the target event.}
    \label{fig:noisy_example}
\end{figure}

\autoref{fig:noisy_example} shows a synthesis that does not meet our success criterion but still reproduces the structure of the target event. The criterion requires a physical deepfake to be indistinguishable from a repeated measurement of the genuine channel. This is the strictest criterion a victim could impose, since such a channel defeats any subsequent processing. Physical events, however, never repeat exactly, and two recordings of the same event differ well beyond the system noise. Thus, we expect that many synthesized channels that miss our criterion still lead the victim to infer the target event. Therefore, the reported rates are only a lower bound on practical attack success.

\section{Physical Deepfakes on \texorpdfstring{\ac{BFI}}{BFI}-based Sensing}
\label{sec:bfi}

\mbox{Wi-Fi} devices report channel estimates as \ac{BFI} to enable beamforming. A growing body of work turns \ac{BFI} into a sensing modality to infer, \eg, human activity~\cite{beamsense}, identity~\cite{todtBFIdIdentityInference2025}, and vital signs~\cite{xiaoLendMeYour2025, kandaBeamformingFeedbackBasedRespiration2025, yiBFMSenseWiFiSensing2024}. In this section, we show that \ac{BFI}-based sensing is vulnerable to physical deepfakes. The attack corresponds to setting S2 in \autoref{fig:observation_settings}, in which a passive attacker overhears the \ac{BFI} reported in the victim environment. As the standard transmits \ac{BFI} in plaintext~\cite{ieee_80211_phy_standard}, the attacker observes the same feedback as the victim sensing system. From this feedback, the attacker observes the target sensing event, learns the metasurface effect, and synthesizes the \ac{BFI} of the target event. Instead of forging feedback frames, the metasurface alters the channel measured by device~$B$, so that the legitimate \ac{BFI} reports carry the physical deepfake.

\Paragraph{Synthesizing Beamforming Feedback}
Sensing on \ac{BFI} operates not on $\mathbf{H}$ directly, but on angles derived from it. Crucially, the attacker observes exactly the \ac{BFI} that the victim sensing system consumes. It can thus learn the effect of its metasurface directly in the feedback domain. We train a forward surrogate to predict the reported feedback for a given configuration. The configuration search of \autoref{sec:config_search} then proceeds unchanged, with the target being a recorded \ac{BFI} stream of the event to be spoofed.

\begin{figure}
    \centering
    \begin{minipage}[c]{0.5\linewidth}
        \centering
        \includegraphics[width=\linewidth]{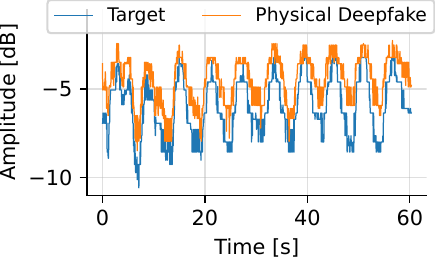}
        \caption{A physical deepfake for \ac{BFI}-based respiration rate estimation.}
        \label{fig:bfi_example}
    \end{minipage}\hfill
    \begin{minipage}[c]{0.48\linewidth}
        \centering
        \footnotesize
        \setlength{\tabcolsep}{2pt}
        \begin{tabular}{@{}lrr@{}}
        \toprule
        \multirow{2}{*}{System} & \multicolumn{2}{c}{Rate [bpm]} \\
        \cmidrule(l){2-3}
         & Target & Deepfake \\
        \midrule
        LeakyBeam~\cite{xiaoLendMeYour2025} & 10.5 & 9.8 \\
        Kanda et al.~\cite{kandaBeamformingFeedbackBasedRespiration2025} & 10.0 & 10.0 \\
        BFMSense~\cite{yiBFMSenseWiFiSensing2024} & 10.2 & 10.2 \\
        \bottomrule
        \end{tabular}
        \captionof{table}{\ac{BFI}-based respiration rate estimation (true respiration at 10.0\,bpm).}
        \label{tab:breathing_rate}
    \end{minipage}
\end{figure}

\Paragraph{Breaking BFI-based Respiration Sensing}
We evaluate the attack against vital-sign sensing, a well-studied \ac{BFI} application. The attacker records the \ac{BFI} of a person breathing at \SI{10.0}{bpm}, synthesizes a configuration sequence that reproduces this feedback, and replays it in the victim environment. \autoref{fig:bfi_example} shows both the target signal and the replayed deepfake which closely tracks the periodic signal variation due to breathing.

We evaluate the deepfake attack with three \ac{BFI}-based respiration sensing systems, LeakyBeam~\cite{xiaoLendMeYour2025}, Kanda~\emph{et~al.}~\cite{kandaBeamformingFeedbackBasedRespiration2025}, and BFMSense~\cite{yiBFMSenseWiFiSensing2024}. Table \ref{tab:breathing_rate} reports the rate each estimator returns for the attacker's genuine target recording and for the replayed deepfake, against the ground-truth rate of \SI{10.0}{bpm}. For all three estimators, the deepfake yields a rate within \SI{0.2}{bpm} of ground truth and closely matches the estimate obtained from the genuine recording, even though the attacker neither knew nor targeted any of these methods. Because the deepfake reproduces the \ac{BFI} itself, every feature these methods derive from it is reproduced as well.

\takeaway{\ac{BFI}-based sensing is insecure: the plaintext feedback that a sensing system consumes also lets a passive attacker make a legitimate device report the feedback of an event that never occurred.}

\section{Synthesis under Constrained Channel Knowledge}
\label{sec:constrained}

\begin{figure}[t]
  \centering
  \includegraphics[width=1.0\columnwidth]{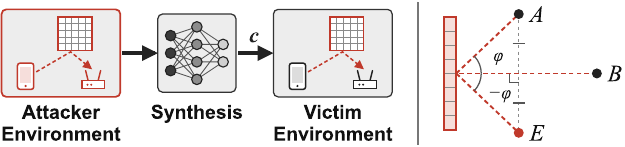}
  \caption{Attacker settings with constrained channel knowledge. Left~(S3): the attacker synthesizes the configurations $\mathbf{c}$ in a physical replica and replays them in the victim environment. Right~(S4): the eavesdropping antenna~$E$ mirrors the victim antenna~$A$ about the symmetry axis through $B$ and the metasurface.}
  \label{fig:constrained_settings}
\end{figure}

In settings S1 and S2, the attacker measures the metasurface's effect on the victim channel directly. This section considers the two settings illustrated in \autoref{fig:constrained_settings}, in which the attacker cannot directly observe the channel but still synthesizes metasurface configurations for physical deepfakes.
In setting S3, the attacker builds a physical replica of the victim sensing system and its geometry, trains on this replica, and then moves the metasurface into the victim environment. In setting S4, the attacker uses an eavesdropping antenna placed inside the victim environment. Our evaluation shows that physical deepfakes remain feasible even under constrained channel knowledge.

We evaluate both settings against the same victim sensing system, a single \mbox{Wi-Fi} link between two commercial off-the-shelf ESP32-C5 devices. The victim extracts \ac{CSI} on channel 112 in the \SI{5}{GHz} band with a bandwidth of \SI{20}{MHz}. We remove the per-packet linear phase caused by sampling time and carrier frequency offsets, following conventional \ac{CSI} preprocessing~\cite{maWiFiSensingChannel2019}. In the following, we first evaluate synthesis quality with the physical replica and the symmetric proxy, and then attack an activity recognition system in both settings.

\subsection{Physical Replica}
\label{sec:constrained:s3}

In S3, the attacker inspects the victim environment and rebuilds the sensing system as a physical replica. As \mbox{Wi-Fi} sensing reuses commodity infrastructure in spaces where people are present, a visitor can note the access point model and its mounting position, purchase the same device, and rebuild the setup with the same geometry.
In the replica, the attacker records sensing events of their own, then installs their metasurface, trains the surrogate model, and synthesizes physical deepfakes of those events, all without access to the victim environment. For the attack, the attacker moves the same metasurface into the victim environment, mounts it at the same position relative to the sensing devices as in the replica, and replays the synthesized configurations. If the replica is sufficiently similar, any event that can be synthesized there can be spoofed in the victim environment. Under the same assumptions, the replica could also be a ray-tracing simulation.

We built such a replica in a different room, with the same device and antenna types and the same geometry of devices and metasurface as the victim setup. We train the surrogate exclusively on measurements from the replica and then move the metasurface to the victim setup. To quantify the similarity, we applied the same sequence of random metasurface configurations to both setups and computed the Pearson correlation between the resulting channel responses. The replica reaches a correlation of 0.73 with the victim setup, against 0.96 between repeated measurements in the victim setup itself.

\begin{figure}[t]
\centering
\begin{subfigure}{0.5\columnwidth}
    \includegraphics[width=\linewidth]{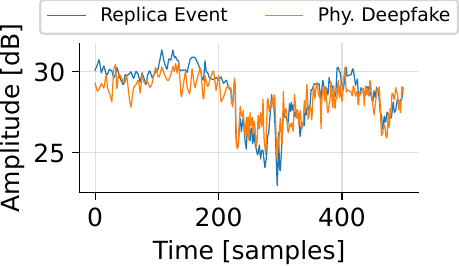}
    \caption{Example of waving}
    \label{fig:clone_replay}
\end{subfigure}\hfill
\begin{subfigure}{0.5\columnwidth}
    \includegraphics[width=\linewidth]{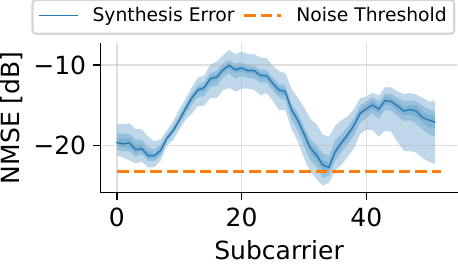}
    \caption{Per subcarrier \ac{NMSE}}
    \label{fig:subc_nmse_clone}
\end{subfigure}
\caption{Results of the attack using a physical replica of the victim setup. (a) An example of a successful physical deepfake of waving. (b) Synthesis error varies with frequency.}
\label{fig:replica_results}
\end{figure}

\autoref{fig:clone_replay} shows a physical deepfake of waving on a single subcarrier at the victim receiver, together with its synthesis target, the attacker's recording in the replica. The physical deepfake transfers between the two setups and reaches measurement-level error in the victim environment.

\autoref{fig:subc_nmse_clone} shows the per-subcarrier \ac{NMSE} between the physical deepfakes and the target events over all 150 recordings. The synthesis error varies across the band. The best subcarriers reach measurement-level synthesis quality, and the average over the full \SI{20}{MHz} band is \SI{-14.7}{dB}.

\subsection{Symmetric Eavesdropping Antenna}
\label{sec:constrained:s4}

Rather than replicating the victim setup, the attacker in S4 places an eavesdropping antenna (the proxy) together with the metasurface inside the victim environment, at the mirror position of the victim antenna, as shown in \autoref{fig:constrained_settings}. This is realistic in public spaces, which are open to anyone. Placing the proxy requires the same physical access as placing the metasurface and thus falls within our threat model. At the proxy, the attacker records the target events in the victim environment and measures the effect of the metasurface. However, the channel at the proxy differs from the channel observed by the victim.

To transfer configurations from the proxy to the victim antenna, the attacker exploits the symmetry of the geometry. The attacker restricts the metasurface to beam-steering configurations, each of which directs the reflected wave toward one azimuth and elevation angle. For each time sample, the attacker selects the beam that best synthesizes the target at the proxy and replays it with inverted azimuth, which yields the corresponding effect at the victim antenna.

\begin{figure}[t]
\centering
\begin{subfigure}{0.5\columnwidth}
    \includegraphics[width=\linewidth]{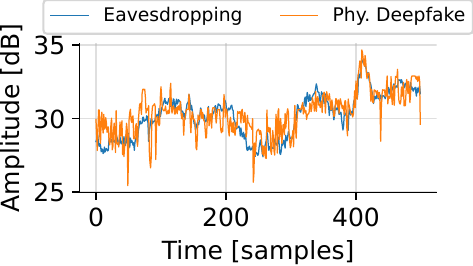}
    \caption{Example of waving}
    \label{fig:proxy_replay}
\end{subfigure}\hfill
\begin{subfigure}{0.5\columnwidth}
    \includegraphics[width=\linewidth]{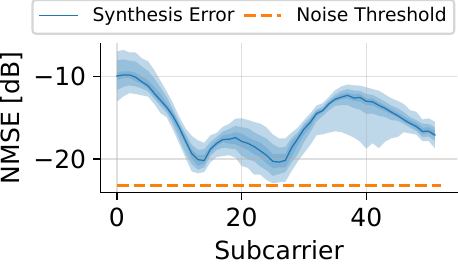}
    \caption{Per subcarrier \ac{NMSE}}
    \label{fig:subc_nmse_proxy}
\end{subfigure}
\caption{Results of the attack using the symmetric eavesdropping antenna. (a) An example physical deepfake of waving. (b) Synthesis error per subcarrier.}
\label{fig:proxy_results}
\end{figure}

\autoref{fig:proxy_replay} shows that a physical deepfake of waving measured at the victim antenna closely follows the target recorded at the proxy. The per-subcarrier error in \autoref{fig:subc_nmse_proxy} follows the same pattern as for the replica, with an average \ac{NMSE} of \SI{-14.5}{dB} across the full band.

\subsection{Attacking Activity Recognition}
\label{sec:constrained:comparison}
We evaluate both the physical replica (S3) and the symmetric proxy (S4) end-to-end against the same classifier, with the synthesis spanning the full \SI{20}{MHz} band. The classifier distinguishes five activities (clapping, jumping, sitting down, walking, and waving), each recorded 30 times at the victim receiver. 
We resample all recordings to a common length so that duration cannot serve as a cue, remove the static path, and classify the resulting Doppler spectrogram with a small 1-D convolutional network.
We use 5-fold cross-validation with 20 recordings per activity for training, 4 for validation, and 6 for testing. \autoref{tab:attack-results} reports the mean and standard deviation over the folds.

The classifier reaches \SI{92.0}{\percent} accuracy on genuine recordings, our reference for both settings. For each attack, we classify three types of inputs. The \emph{targets} are genuine recordings that the attacker captures of the events to reproduce. The \emph{physical deepfakes} are measured at the victim receiver while the metasurface replays the synthesized configurations. The \emph{shuffled} deepfakes have their samples randomly permuted in time, which destroys the motion pattern but preserves the amplitude statistics. At \SI{19.3}{\percent} and \SI{20.7}{\percent}, this baseline coincides with the chance level of \SI{20}{\percent}.

The two settings differ in how the attacker obtains targets. In the proxy setting, the attacker records the same events as the victim, and the targets reach \SI{90.7}{\percent}, on par with the victim's own recordings, as the Doppler signature of the motion is preserved at the proxy. The physical deepfakes reach \SI{58.0}{\percent}, three times the shuffled baseline. In the replica setting, the attacker relies on their own repetitions of the activities, which limits the targets to \SI{46.7}{\percent}. The physical deepfakes retain most of this accuracy at \SI{34.0}{\percent}.

\begin{figure}[t]
\centering
\begin{subfigure}{0.325\columnwidth}
    \includegraphics[width=\linewidth]{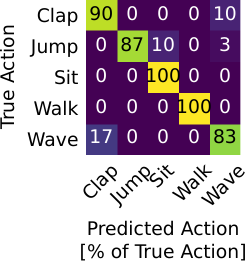}
    \caption{Genuine}
    \label{fig:confusion_victim}
\end{subfigure}\hfill
\begin{subfigure}{0.325\columnwidth}
    \includegraphics[width=\linewidth]{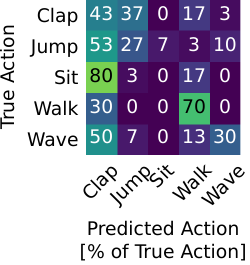}
    \caption{Replica (S3)}
    \label{fig:confusion_replica}
\end{subfigure}\hfill
\begin{subfigure}{0.325\columnwidth}
    \includegraphics[width=\linewidth]{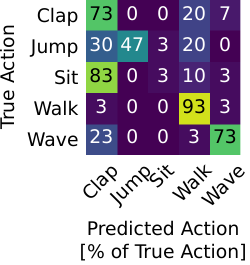}
    \caption{Proxy (S4)}
    \label{fig:confusion_proxy}
\end{subfigure}
\caption{Confusion matrices of the activity classifier for (a) genuine test recordings and physical deepfakes from (b) the physical replica and (c) the symmetric proxy.}
\label{fig:confusion}
\end{figure}

\autoref{fig:confusion} shows that the misclassified deepfakes concentrate on clapping and walking in both settings. We attribute this to the synthesis error, which varies from packet to packet and therefore appears as broadband noise in the Doppler spectrogram. This noise masks slow activities such as sitting down, which carry little energy at high Doppler frequencies, and biases the classifier towards walking, as also observed for the shuffled baseline.

\takeaway{Physical deepfakes deceive activity recognition even when the attacker cannot observe the victim channel, using either a physical replica of the victim setup or an eavesdropping antenna.}

\begin{table}[t]
  \centering
  \caption{Accuracy of the victim's classifier on genuine recordings and on the targets, physical deepfakes, and time-shuffled deepfakes of the replica (S3) and proxy (S4) attacks (mean $\pm$ standard deviation over five folds).}
  \label{tab:attack-results}
  \begin{tabular}{lll}
    \toprule
    Scenario & Input & Accuracy [\%] \\
    \midrule
    Victim                        & Genuine & $92.0 \pm 5.6$ \\
    \midrule
    \multirow{3}{*}{Replica (S3)} & Targets  & $46.7 \pm 14.3$ \\
                                  & Physical Deepfake & $34.0 \pm 13.4$ \\
                                  & Shuffled & $20.7 \pm 4.3$ \\
    \midrule
    \multirow{3}{*}{Eavesdropping (S4)}   & Targets  & $90.7 \pm 2.8$ \\
                                  & Physical Deepfake & $58.0 \pm 9.6$ \\
                                  & Shuffled & $19.3 \pm 2.8$ \\
    \bottomrule
  \end{tabular}
\end{table}

\section{Environment Cloning}
\label{sec:cloning}
\begin{figure}[t]
  \centering
  \includegraphics[width=1.0\columnwidth]{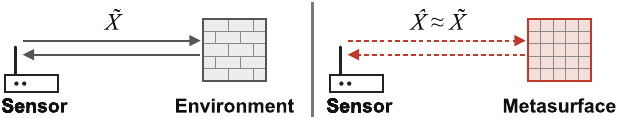}
  \caption{Environment cloning~(S5). Left: the sensor observes the reflection of a part of the environment, the target event $\tilde{\mathbf{X}}$. Right: the attacker replaces this part with a metasurface that mimics its reflection, so the sensor infers $\hat{\mathbf{X}} \approx \tilde{\mathbf{X}}$.}
  \label{fig:environment_cloning}
\end{figure}

Our final evaluation addresses setting S5 in \autoref{tab:attacker_settings}, in which the attacker has one-time initial access to the victim environment. During this visit, the attacker profiles the reflection behavior of a part of the environment, \eg, an object, but obtains no knowledge of the sensing system. S5 thus departs from the assumption in \autoref{sec:system_model} that the attacker knows the geometry of the victim setup. Instead of targeting the sensing system, the attacker therefore clones this part of the environment, as shown in \autoref{fig:environment_cloning}. In their own lab, the attacker synthesizes a metasurface configuration whose response $\hat{\mathbf{X}}$ approximates the profiled reflection characteristics $\tilde{\mathbf{X}}$. When returning to carry out the attack, the attacker places the configured metasurface as a stand-in for the target or moves it to imitate the target's motion.

We evaluate the attack with a transmit antenna directed at the object and a receive array of seven adjacent antennas. As the attacker in S5 does not know where the sensing system is located, we require the clone to reproduce the object's reflection at all seven antennas simultaneously. The configuration thus approximates the object's scattered field across the angular sector covered by the array, and the seven antennas act as stand-ins for possible receiver positions.

\begin{figure}
    \centering
    \includegraphics[width=0.95\linewidth]{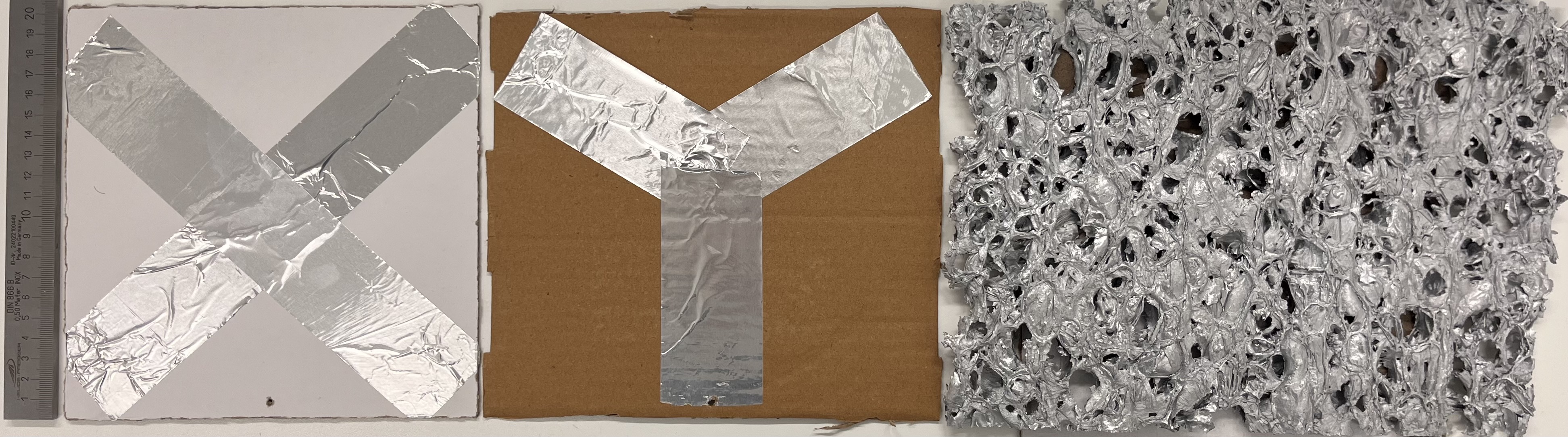}
    \caption{Objects used for the cloning experiments. Targets~1--3 from left to right: cardboard sheets with metallic tape in the shape of the letters X and Y, and a fully metallic object with an irregular surface structure.}
    \label{fig:clone_obj}
\end{figure}

We measure the channel with the \ac{VNA} from \autoref{sec:setup} at \SI{5.56}{GHz}, using a subcarrier spacing of \SI{312.5}{kHz} over a bandwidth of \SI{20}{MHz} to match standard \mbox{Wi-Fi}. As targets, we use the three objects shown in~\autoref{fig:clone_obj}: two cardboard sheets carrying metallic tape in the shape of the letters X (Target~1) and Y (Target~2), and one fully metallic object with an irregular surface structure, a complex reflector (Target~3). Targets~1 and~2 serve as a pair of highly similar objects. For each target, we record a static reference measurement $\tilde{\mathbf{X}} \in \mathbb{C}^{7\times57}$.

As in the previous settings, we synthesize the deepfake using a surrogate model of the metasurface's effect on the channel. In contrast to settings S1--S4, where the metasurface is located close to the receiver, it now takes the position of the object it is meant to replace. For each reference $\tilde{\mathbf{X}}$, we then use the surrogate model to find a metasurface configuration whose response $\hat{\mathbf{X}}$ approximates it.

\begin{figure}
    \centering
    \includegraphics[width=0.9\linewidth]{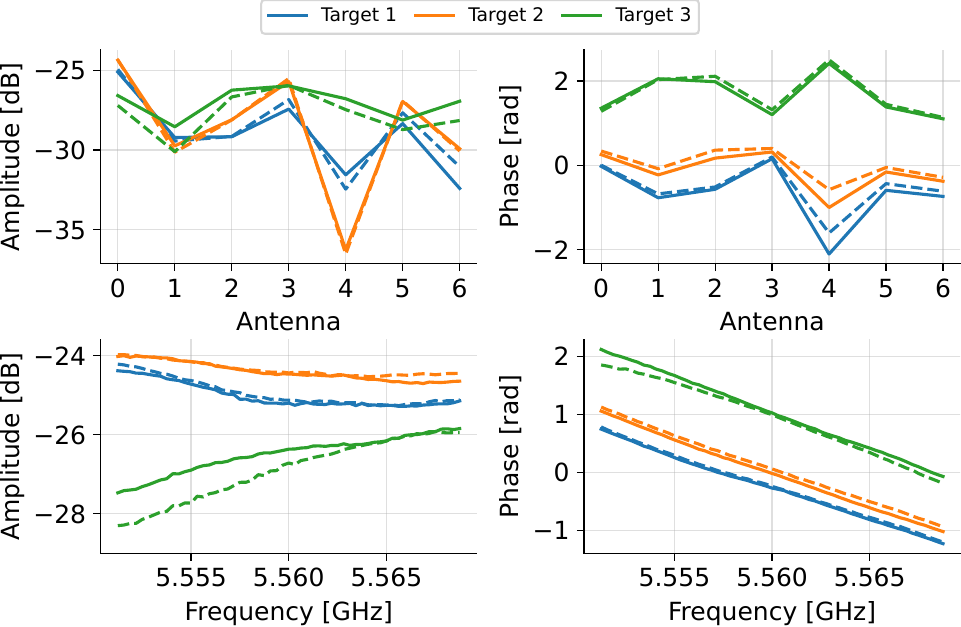}
    \caption{Object cloning results for Targets~1--3, comparing the measured reference $\tilde{\mathbf{X}}$ of each target (solid) with the synthesized deepfake $\hat{\mathbf{X}}$ (dashed). Columns show amplitude (left) and phase (right), while rows show the variation across the seven receive antennas at \SI{5.56}{GHz} (top) and across frequency at antenna~0 (bottom).}
    \label{fig:cloning_results}
\end{figure}

\Paragraph{Results}
\autoref{fig:cloning_results} compares the references $\tilde{\mathbf{X}}$ with the synthesized deepfakes $\hat{\mathbf{X}}$ across the receive antennas and across frequency. For all three targets, the deepfake closely follows its reference in both amplitude and phase. Over all antennas and subcarriers, the deepfakes reach an \ac{NMSE} of \SI{-17.2}{dB}, \SI{-19.1}{dB}, and \SI{-15.1}{dB} for Targets~1,~2, and~3. This is well below the smallest \ac{NMSE} between two different targets, \SI{-11.2}{dB} between Targets~1 and~2. Each deepfake is thus at least \SI{3.9}{dB} closer to its target than even the most similar other object.

\Paragraph{Robustness to Displacement}
The previous experiment evaluates the clone at a single, fixed position. In practice, the attacker cannot place the cloned object at the exact location used for synthesis, and the sensing system may observe it over a range of positions. To test whether a synthesized configuration holds under such displacement, we move the objects and the metasurface with a linear actuator across 89 positions spaced \SI{0.5}{cm} apart. The metasurface retains the configuration synthesized at the first position. At each position, we measure the resulting response and compute its \ac{NMSE} against the corresponding target reference. 

\begin{figure}
    \centering
    \includegraphics[width=0.7\linewidth]{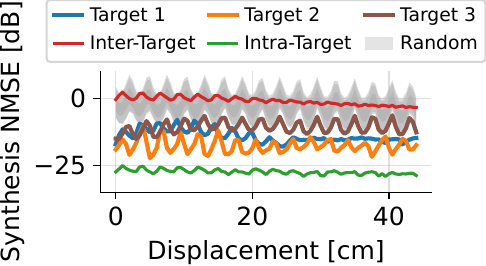}
    \caption{Synthesis error for environment clones as the metasurface is moved along the linear actuator.}
    \label{fig:cloning_displacement}
\end{figure}

\autoref{fig:cloning_displacement} shows the result, together with the \ac{NMSE} of random configurations and the mean inter- and intra-target \ac{NMSE} for reference. Across the entire range, every deepfake stays below both the random configurations and the inter-target \ac{NMSE}, \ie, at any position, the clone is closer to its own target than any two targets are to each other.
All curves oscillate with a period of roughly \SI{2.7}{cm}, half the wavelength, as the reference power oscillates with the interference of direct and reflected paths while the absolute synthesis error stays constant. The synthesis and intra-target curves show no systematic trend over the range, whereas the inter-target \ac{NMSE} decreases with displacement. We attribute the latter to the growing dominance of the static path over the object's reflection as the object moves away from the array, which makes the targets increasingly similar to one another, independent of the clone. Overall, the deepfake remains tied to its target across all positions.

\takeaway{A metasurface clones the reflection of physical objects without any knowledge of the sensing system, and the clone remains closer to its target than to any other object when displaced.}

\section{Discussion and Future Work}
\label{sec:discussion}

\subsection{Experimental Setup and Results}

To record training samples, the attacker cycles the metasurface through its configuration space, which perturbs the victim channel and might be noticed. 
However, random configurations resemble channel noise rather than a plausible physical event. The attacker can further sample an ordered sequence of configurations that differ in only a few elements, keeping the channel variation small and gradual.
In S3 and S5, the attacker samples outside the victim environment altogether.

The attacker places the metasurface in the vicinity of the sensing devices, as the metasurface path attenuates with the distances to both $A$ and $B$. A larger aperture compensates for distance, and the attacker only needs to induce channel variation on the order of what a human reflector causes.

\subsection{Limitations}

Our attack assumes that the environment is static between the acquisition of channel knowledge and the attack. Changes in the environment, \eg, moving devices or people, render the synthesized metasurface configurations outdated, so that the attacker has to update their channel knowledge. However, in settings S1 and S2, the attacker observes the victim channel continuously and can update their channel knowledge at any time.
Moreover, the attacker requires physical access to place the metasurface in the victim environment. The attack is further bound to the operating frequency bands of the metasurface and cannot synthesize channels on frequencies beyond them. Further, we only synthesize events occurring outside the \ac{LOS} between $A$ and $B$, matching the sensing of reflective events. Still, events occurring within the \ac{LOS} remain synthesizable as long as the metasurface induces channel variation of the required magnitude. Finally, a physical deepfake reproduces a recording of the target event, which the attacker has to obtain beforehand. We argue that an exact replica of the target measurement is the strongest variant of the attack, but it is not strictly needed for attack success as wireless sensing has to be inherently domain-invariant, because radio propagation encodes sensed events differently across environments.

\subsection{Countermeasures}
Countermeasures can deny the attacker the channel knowledge listed in \autoref{tab:attacker_settings}, so that they cannot determine suitable metasurface configurations. S1 is mitigated by ensuring that the sensing partner is honest. S2 can be thwarted by protecting the \ac{BFI}, \eg, through encryption or randomized precoding during channel sounding~\cite{xiaoLendMeYour2025}.
In S3 and S4, the attacker transfers channel knowledge from a replica or an eavesdropping antenna to the victim channel. Confidential sensor configurations, \eg, randomized beamforming, prevent the attacker from replicating the victim setup. In S5, the attacker creates an electromagnetic clone of the victim environment and thus needs no channel knowledge at all. These countermeasures are setting-specific and give no guarantees against attackers who find other ways to obtain channel knowledge. Two general countermeasures apply regardless of the setting.
First, concealing the physical environment prevents the attacker from placing a metasurface at all, and thus it yields a physically confined wireless channel that is known to be benign.
Second, the victim can make synthesis harder by sensing across more \ac{DoF} with consistency checks across links, by lowering the system noise, and by frequency hopping beyond the bandwidth of the metasurface.

\subsection{Future Work}

We synthesize the metasurface configuration for each time instant independently. Future work could synthesize entire configuration sequences to ensure temporal smoothness. Moreover, configuration sequences could be upsampled with gradual fading between consecutive configurations. 
Furthermore, we use a single metasurface to synthesize all \ac{DoF} at once. Since synthesis quality drops as the number of jointly synthesized \ac{DoF} grows, multiple metasurfaces could share the load, \eg, with a dedicated metasurface per frequency range or spatial link.
Finally, physical deepfakes could extend to other sensing domains in which metasurfaces exist, such as acoustic~\cite{tianProgrammableAcousticMetasurfaces2019} and optical~\cite{shaltoutSpatiotemporalLightControl2019} sensing.

\section{Related Work}

\Paragraph{Metasurface Security Applications}
The security literature investigates metasurfaces for both benign and malicious purposes. Metasurfaces have been used for manipulation of wireless sensing at the feature level, both in benign~\cite{staatIRShieldCountermeasureAdversarial2022, wangSpreadSpectrumSelectiveCamouflaging2021} and malicious~\cite{jiangRISirenWirelessSensing2024, zhouRIStealthPracticalCovert2023} applications. Recent analytical studies show that metasurfaces can spoof the angle-of-arrival estimates of monostatic radar~\cite{gavrasRISEnabledSpoofingAdversary2026} as well as Doppler and angle-of-departure estimates in vehicular \ac{ISAC} networks~\cite{shuiAnalysisDetectionRISbased2025a}. In the context of wireless communication, metasurfaces have been used to defend against or mount jamming attacks~\cite{yangIntelligentReflectingSurface2021, staatMirrorMirrorWall2022, mackensenSpatialDomainWirelessJamming2025} and to thwart or assist eavesdropping~\cite{weiMetasurfaceenabledSmartWireless2023, liProtegoSecuringWireless2022, shaikhanovMetasurfaceintheMiddleAttackTheory2022}. Metasurfaces can also randomize wireless channels~\cite{staatIntelligentReflectingSurfaceAssisted2021}, for instance, to support the authentication of transmitters~\cite{tomasinChallengeResponsePhysicalLayer2022} and the verification of their location~\cite{zhangRISCLARevivingCSIBased2026}.

\Paragraph{Manipulation of Wireless Sensing}
Physical manipulation of wireless sensing originated from the defense sector with the goal of deceiving adversarial radar, \eg, through decoys~\cite{griffthsDdayDeceptionOperations2015}, signal replay~\cite{roomeDigitalRadioFrequency1990}, and malicious reflection~\cite{kozlovRadarRangeDeception2023}. As wireless sensing is used in many civilian applications today, manipulation has become a civilian threat as well. In automotive radar, adversaries can conceal real objects or inject false ones, either with engineered reflectors~\cite{chenMetaWaveAttackingMmWave2023, zhuTileMaskPassiveReflectionbasedAttack2023} or by re-radiating signals~\cite{reddyvennamMmSpoofResilientSpoofing2023, komissarovSpoofingAttacksVehicular2021}. Similarly, sensing of humans and indoor environments can be manipulated with crafted objects and smart surfaces to fake vital signs~\cite{guoAttackingMmWaveenabledChest2026}, bypass intrusion detection~\cite{zhouRIStealthPracticalCovert2023, akinwalePracticalReflectionManipulation2025}, simulate falls~\cite{jiangRISirenWirelessSensing2024}, and deceive indoor localization~\cite{hanLokiPhysicalWorldAdversarial2025}. Physical objects can also fool camera-lidar sensing in autonomous vehicles~\cite{caoInvisibleBothCamera2021, zhuMaliciousAttacksMultiSensor2024}, X-ray imaging~\cite{liuXAdvPhysicalAdversarial2023}, and satellite radar~\cite{liBroadbandInformationMetasurfaceAssisted2024}.

\Paragraph{Defenses against Sensor Spoofing}
Sensors in general are susceptible to manipulated inputs~\cite{fuRisksTrustingPhysics2018, zhangDolphinAttackInaudibleVoice2017, kuneGhostTalkMitigating2013, sonRockingDronesIntentional2015}, as they implicitly assume that the physical signals they capture are authentic. Countermeasures typically rely on anomaly detection, both in wireless sensing~\cite{gengMmGuardCountermeasurePhysical2026, rothAnomalyDetectionSensing2026} and beyond~\cite{parkSensorAttackDetection2015}. Yet, such defenses presuppose that attacks produce detectable anomalies~\cite{yanSoKMinimalistApproach2020, giraldoSurveyPhysicsBasedAttack2019}. Consequently, they make attacks harder but are inherently defeated by an adversary whose manipulation is consistent with the expected physical observation~\cite{urbinaLimitingImpactStealthy2016}.

\section{Conclusion}

In this work, we presented physical deepfake attacks against wireless sensing systems. We formalized the attack and realized it with metasurfaces that manipulate the physical sensing environment. This novel class of attacks accurately reproduces the wireless channel of genuine physical events. From a security perspective, our attack is thus the channel-level equivalent of classical message-level replay attacks. We showed that the attack spoofs wireless sensing systems independent of the signal features they use for inference. Our work exposes the vulnerability of wireless sensing to physical manipulation, as sensing inherently trusts its physical observations of the radio environment, which is potentially adversary-controlled. Given the global efforts to implement \ac{ISAC}, we argue that wireless sensing for safety-critical applications needs additional security measures.

\ifCLASSOPTIONcompsoc
  \section*{Acknowledgments}  
\else
  \section*{Acknowledgment}
\fi

We thank NovoFlect GmbH, Simon Tewes, Markus Heinrichs, and Rainer Kronberger for providing the metasurface prototypes.
Furthermore, we thank Christian Zenger for his guidance as the first author's doctoral advisor, and Ilyes Lajili and Ali Kalife for technical support.
This work was in part funded by the Deutsche Forschungsgemeinschaft (DFG, German Research Foundation) under Germany’s Excellence Strategy - EXC 2092 CaSa - 390781972.

\bibliographystyle{IEEEtran}
\bibliography{phydeep_refs}

\appendices

\section{Forward Surrogate and Configuration Search}
\label{app:surrogate}

\Paragraph{Surrogate Architecture}
The forward surrogate $\hat{\mathbf{H}}(\mathbf{c})$ (\autoref{sec:framework}) maps the $L = 256$ binary element states to the channel $\hat{\mathbf{H}}(\mathbf{c}) \in \mathbb{C}^{s \times k}$. Its prediction is the sum of two parallel branches. The linear branch maps the element states directly to the channel. The convolutional branch views the configuration as a $16\times16$ aperture image. A $3\times3$ convolution first maps the single-channel image to $256$ feature channels. Six residual blocks then process the resulting feature map, each applying two $3\times3$ convolutions with GroupNorm and GELU. Each spatial link is assigned a learned query vector that attends over the $256$ spatial feature vectors, producing one vector per link. A shared decoder network with GELU activation maps each link vector to the link's row of $\hat{\mathbf{H}}(\mathbf{c})$, \ie, its channel across the $k$ subcarriers. The architecture has about $14.7$\,million parameters.

\Paragraph{Surrogate Training}
We standardize each of the $2 \times s \times k$ real-valued outputs and minimize the mean-squared error between the predicted and measured channels using Adam~\cite{kingma2015adam}. Training uses a learning rate of $3 \times 10^{-4}$, which is halved each time the validation loss plateaus. For fine-tuning to a new environment, we follow~\cite{kumar2022finetuning}: we first fit a linear read-out layer that maps the outputs of the existing surrogate to the new environment, and then fine-tune the entire architecture with a learning rate of $10^{-5}$.

\Paragraph{Configuration Search}
We run the gradient-based configuration search (\autoref{sec:config_search}) with $32$ restarts, each running Adam for $1000$ iterations at a learning rate of $0.2$. One
restart is a warm start, a configuration in $[0,1]^{256}$ that we optimize on
the surrogate's linear branch for $400$ steps and then map to logits, and the other
$31$ restarts are random logits
$\mathbf{z} \sim \mathcal{N}(\mathbf{0}, \mathbf{I}_{256})$. In each iteration, we binarize the sigmoid outputs to a realizable configuration with a threshold of $0.5$, evaluate the forward surrogate $\hat{\mathbf{H}}(\mathbf{c})$ on it, and backpropagate the  error in the prediction to the logits with a straight-through estimator. Each restart keeps the best
configuration it visits.

From the $32$ restarts of each frame, we keep the $4$ with the lowest error and refine them in two steps. First, we apply single-element flips to each of them, testing each of the $256$ elements in turn and accepting any flip that lowers the error. We loop over the $256$ elements up to $4$ times and stop early once a sweep accepts no bit
flip. Next, we apply pair flips to the best of the $4$ configurations by flipping two elements together. Instead of testing all pairs, we predict each pair's channel change as the sum of its two single-flip changes and evaluate only the $128$ pairs with the lowest predicted error. We repeat the pair step up to $60$ times and stop once no single or pair move lowers the error. Together, the single and pair flips reduce the deepfake error by a few tenths of a dB.

On a single NVIDIA H100 GPU, we design 50 frames with 32 restarts each in a single batch, which takes about \SI{3.5}{s} per frame. The attacker may design the configuration sequence offline before replaying it on the hardware.

\end{document}